\documentclass[11pt,a4paper]{article}

\usepackage{times,epsfig,wrapfig}

\advance \textwidth by -5cm
\advance \textheight by -6cm
\def\Vec#1{\mbox{\boldmath $ #1 $}}

\def\e{\begin{equation}}
\def\f{\end{equation}}
\def\_#1{{\bf #1}}

\def\.{\cdot}

\begin{document}

\title{\bfseries{Landau-Lifshitz theory of \\ single susceptibility Maxwell equations}}

\def\affil#1{\begin{itemize} \item[] #1 \end{itemize}}

\author{\bfseries K. Cho}



\date{}

\maketitle
\thispagestyle{empty}

\affil{Institute of Laser Engineering, Osaka University, \\
Suita, Yamada-oka 2-6, Postal Code 565-0871, Japan \\
email: k-cho@kcc.zaq.ne.jp}

\begin{abstract}
\noindent 
 The conflicting arguments given in the discussion forum of Metamaterials 2011 
on the possible forms of macroscopic Maxwell equations are lead to a convergence 
by noting the relationship among the employed material variables for each scheme.  
The three schemes by Chipouline et al. using (A) standard $\Vec{P}$ and $\Vec{M}$ 
(Casimir form), (B) generalized electric polarization $\Vec{P}_{LL}$ 
(Landau-Lifshitz form), (C) generalized magnetic polarization $\Vec{M}_{A}$ 
(Anapole form) are compared with (D) the present author's scheme using standard 
current density $\Vec{J}$. From the reversible relations among the transverse 
components of these vectors, one can easily rewrite one scheme into another.  
The scheme (D), the only one among the four providing the first-principles 
expressions of susceptibility and 
also leading to a non-phenomenological Casimir form in terms of the four generalized 
susceptibilities between $\{\Vec{P},\Vec{M}\}$ and $\{\Vec{E},\Vec{B}\}$, is 
concluded to be a more natural form than (B) and (C) as a single susceptibility 
theory. 
\end{abstract}

\subsection*{1.~Introduction}
In the conventional macroscopic Maxwell equations (M-eqs), the variables of matter are 
usually represented by the electric and magnetic polarizations $\Vec{P}$ and $\Vec{M}$, 
in contrast to the microscopic M-eqs where we need only current (and charge) density 
$\Vec{J}$ ($\rho$).  In view of the general relationship $\Vec{J} = \partial \Vec{P}/
\partial t + \nabla \times \Vec{M}$, the description in terms of $\Vec{P}$ and $\Vec{M}$ 
uses redundant variables.  The constitutive equation to be required should be a single 
equation relating a single vector quantity of matter with that of EM field ($\Vec{E}$ 
or $\Vec{B}$, or else). This aspect has been known for a long time, but not 
satisfactorily worked out.  Landau and Lifshitz (LL) \cite{LL} proposed to use the M-eqs 
$\nabla \times \Vec{E} = - \partial \Vec{B}/\partial t, \ \ \nabla \times \Vec{B} 
= \partial \Vec{D}/\partial t$, (i.e., $\Vec{H} = \Vec{B}$).  
This means that one uses a new variable $\Vec{P}_{LL}$,  defined by  
$\Vec{J} = \partial \Vec{P}_{LL}/\partial t$, containing the both characters 
of $\Vec{P}$ and $\Vec{M}$ \cite{IK, AG}.   The constitutive equation 
in this case relates $\Vec{D}$ and $\Vec{E}$ through a single susceptibility. 
LL discuss its symmetry properties, but not its quantum mechanical expression. \\

  Recently, Chipouline et al. (CST) \cite{CST} considered the possible forms of 
macroscopic M-eqs obtained by macroscopic averaging of microscopic M-eqs.   
Noting the non-uniqueness in the equations $\nabla \cdot \Vec{P} = - \rho$ and 
$\nabla \cdot \Vec{J} = -\partial \rho/\partial t$, i.e., the fact that $\Vec{P}$ 
and $\Vec{J}$ may contain $\nabla \times$ of an arbitrary vector function, 
they consider three choices of matter variables, (A) usual $\Vec{P}$ and 
$\Vec{M}$ (Casimir form), (B) $\Vec{P}_{LL}$ (LL form), (C) $\Vec{M}_{A}$ 
defined by $\Vec{J} = \nabla \times \Vec{M}_{A}$ (Anapole form).  They discuss 
the relationship among the three cases, including the possibility of their 
mutual transformation, but no quantum mechanical consideration is given 
about the form of susceptibilities. \\

Another single susceptibility theory for macroscopic description was developed 
by the present author in terms of current density $\Vec{J}$ \cite{Cho}. 
This theory leads to the constitutive equation 
\begin{equation}
  \Vec{J}(\Vec{k}, \omega) = \chi_{\rm em}(\Vec{k}, \omega) \cdot 
  [\Vec{A}(\Vec{k}, \omega) + (c/i\omega) \Vec{E}_{\rm extL}(\Vec{k}, \omega)]
\end{equation} 
where $\Vec{A}$ is the transverse (T) vector potential in Coulomb gauge, and 
$\Vec{E}_{\rm extL}$ 
the longitudinal (L) electric field due to external charge density. (The L 
field due to internal charge density is taken into account as the Coulomb 
potential in matter Hamiltonian.) The macroscopic susceptibility $\chi_{\rm em}$ 
is derived via long wavelength approximation of the microscopic (nonlocal) one as 
\begin{equation}
 \chi_{\rm em}(\Vec{k},\omega) = V \sum_{\nu} \big[ \bar{g}_{\nu}(\omega) 
               \tilde{\Vec{I}}_{0\nu}(\Vec{k})\tilde{\Vec{I}}_{\nu 0}(-\Vec{k}) 
        + \bar{h}_{\nu}(\omega) 
               \tilde{\Vec{I}}_{\nu 0}(\Vec{k})\tilde{\Vec{I}}_{0\nu}(-\Vec{k}) \big] \ ,
\end{equation} 
where $\nu$ (and $\mu$ below) is the quantum number of matter eigenstates, 
$V$ the quantization volume of $\Vec{k}$, 
 $\bar{g}_{\nu}(\omega) = 1/(E_{\nu 0} - \hbar \omega - i0^+) - 1/E_{\nu 0}, $  
 $\bar{h}_{\nu}(\omega) = 1/(E_{\nu 0} + \hbar \omega + i0^+) - 1/E_{\nu 0}, $
and $E_{\nu 0}$ the excitation energy from the ground state.  The matrix element 
of current density can be written as 
\begin{equation}
\label{eqn:CD-LWA2}
 \tilde{\Vec{I}}_{\mu\nu}(\Vec{k}) = (\exp(-i\Vec{k}\cdot\bar{\Vec{r}})/V) 
         \big[\bar{\Vec{J}}_{\mu\nu} - i \Vec{k} \cdot\bar{\bf Q}_{\mu\nu}^{(\rm e2)} 
              + i \Vec{k} \times \bar{\Vec{M}}_{\mu\nu} + O(k^2) \big] 
\end{equation}
\begin{equation}
  \bar{\Vec{J}}_{\mu\nu} = \int{\rm d}\Vec{r}\ \langle\mu|\Vec{J}_{0}|\nu\rangle, \ \ 
  \bar{\Vec{M}}_{\mu\nu} = \bar{\Vec{M}}_{\mu\nu}^{(\rm spin)} 
                           + \bar{\Vec{M}}_{\mu\nu}^{(\rm orb)}\ ,
\end{equation}
\begin{eqnarray}
\label{eqn:MQE2}
 \Vec{k}\cdot\bar{\Vec{Q}}_{\mu\nu}^{(\rm e2)} =  
       \sum_{\ell} \frac{e_{\ell}}{2m_{\ell}}\ \int{\rm d}\Vec{r}\ 
 \{ <\mu|  (\Vec{r}_{\ell} - \bar{\Vec{r}}) \Vec{k} \cdot \Vec{p}_{\ell}\ \delta(\Vec{r}_{\ell} - \Vec{r}) 
  + \delta(\Vec{r}_{\ell} - \Vec{r})\ (\Vec{r}_{\ell} - \bar{\Vec{r}}) \Vec{k} \cdot\Vec{p}_{\ell} |\nu> \}\ , 
\end{eqnarray}
\begin{equation}
\label{eqn:M1orb}
 \bar{\Vec{M}}_{\mu\nu}^{(\rm orb)} 
          = \sum_{\ell} \frac{e_{\ell}}{2m_{\ell}}\ \int{\rm d}\Vec{r}\  
    <\mu|  \Vec{L}_{\ell}(\bar{\Vec{r}})\ \delta(\Vec{r}_{\ell} - \Vec{r}) 
   + \delta(\Vec{r}_{\ell} - \Vec{r})\ \Vec{L}_{\ell}(\bar{\Vec{r}}) |\nu> \ , 
\end{equation}
where $\Vec{J}_{0}(\Vec{r}) = \sum_{\ell} (e_{\ell}/2m_{\ell}) 
\{ \Vec{p}_{\ell}\ \delta(\Vec{r}_{\ell} - \Vec{r}) + 
\delta(\Vec{r}_{\ell} - \Vec{r}) \Vec{p}_{\ell}\}$, and 
$\Vec{L}_{\ell}(\bar{\Vec{r}}) = (\Vec{r}_{\ell} - \bar{\Vec{r}}) \times \Vec{p}_{\ell}$, 
is the angular momentum of the $\ell$-th particle with respect to the center coordinate 
$\bar{\Vec{r}}$ of the $(\mu, \nu)$ transition  
to make Taylor expansion of $\tilde{\Vec{I}}_{\mu\nu}(\Vec{k})$.
The zero-th and first order moments 
$\bar{\Vec{J}}_{\mu\nu}, \bar{\bf Q}^{(\rm e2)}_{\mu\nu}, 
\bar{\Vec{M}}_{\mu\nu}$ are nonzero for electric dipole, electric quadrupole and 
magnetic dipole transitions, respectively.  This result covers all the cases of 
linear response, including chiral susceptibility \cite{Cho,Cho-MM11}. 
This scheme should be added to the list of CST as the fourth item (D) 
specified by the use of matter variable $\Vec{J}$.  It may be called 
"natural form", since it does not use unfamiliar variables as in (B) and (C). \\

A discussion forum was held in the Metamaterials 2011 (Barcelona) about this 
problem.  The large number of participants shows the general interests 
in this very fundamental problem.  The discussions were rather conflicting  
with premature arguments, since the participants had not been well informed 
beforehand about the contents of other parties.  Later the present author 
made a visit to have more detailed discussions with the CST group, which 
has resulted in this article unifying the schemes (A, B, C) with (D). 

\subsection*{2.~Unification of the four forms}

The essential point for the unification of different schemes is to note 
that the non-uniqueness introduced by $\nabla \cdot \Vec{P} = - \rho$ and 
$\nabla \cdot \Vec{J} = -\partial \rho/\partial t$ is only for the T  
components of $\Vec{P}$ and $\Vec{M}$, which is because the two equations 
give constraint only to the L components.  This requires a refinement in 
the defining equations of the CST's classification, i.e, instead of 
$\Vec{J} = \partial \Vec{P}_{LL}/\partial t = \nabla \times \Vec{M}_{A}$
we should use 
\begin{equation} 
\label{eqn:def-eq}
 \Vec{J}^{(T)} = \partial \Vec{P}_{LL}^{(T)}/\partial t \ \ 
               = \nabla \times \Vec{M}_{A}^{(T)} \ . 
\end{equation}
This means that the L component of $\Vec{J}$ is common to all the 
schemes (A, B, C, D).  Namely, the choice of the variables should be 
\begin{itemize}
\item [(A)]  Casimir form : $\Vec{P}^{(T)},\ \Vec{M}^{(T)}$ and $\Vec{J}^{(L)}$ \ ,
\item [(B)]  LL form : $\Vec{P}_{LL}^{(T)}$ and $\Vec{J}^{(L)}$\ ,
\item [(C)]  Anapole form : $\Vec{M}_{A}^{(T)}$ and $\Vec{J}^{(L)}$\ ,
\item [(D)]  Natural form : $\Vec{J}^{(T)}$ and $\Vec{J}^{(L)}$\ .
\end{itemize}
In (A) and (B), $\Vec{J}^{(L)}$ may be replaced by $\Vec{P}^{(L)}$ 
and $\Vec{P}_{LL}^{(L)}$, respectively, which are equivalent to 
$(i/\omega) \Vec{J}^{(L)}$. \\

Equation (\ref{eqn:def-eq}) can be solved as 
\begin{equation} 
  \Vec{P}_{LL}^{(T)}(\Vec{k}, \omega) 
            = (i/\omega)\ \Vec{J}^{(T)}(\Vec{k}, \omega), \ \ \ 
  \Vec{M}_{A}^{(T)}(\Vec{k}, \omega) 
            = (i/k^2) \Vec{k} \times \Vec{J}^{(T)}(\Vec{k}, \omega). 
\end{equation}
This means that the constitutive equation for $\Vec{J}$, already known 
in the scheme (D), can be transformed into those for $\Vec{P}_{LL}$ 
and $\Vec{M}_{A}$ as  
\begin{eqnarray}
 \Vec{P}_{LL}^{(T)} &=& (i/\omega) \chi_{\rm em} \cdot 
                  [\Vec{A} - (i/\omega) \Vec{E}_{\rm extL}] \ , \\
 \Vec{M}_{A}^{(T)} &=& (i/k^2) \Vec{k} \times 
                 [\chi_{\rm em}\cdot \{\Vec{A} - (i/\omega) \Vec{E}_{\rm extL}\}]^{(T)} \ .
\end{eqnarray}

The transformation from (D) to (A) is discussed in \cite{Cho} (Chap.3) 
and \cite{Cho-MM11} by using the explicit expression of $\chi_{\rm em}$, which 
defines the four susceptibilities, $\{(\chi_{\rm eE}, \chi_{\rm eB}), 
(\chi_{\rm mE}, \chi_{\rm mB})\}$, i.e., the electric and magnetic 
susceptibilities induced by $\Vec{E}$ and $\Vec{B}$.  This rewriting is 
reversible if one uses the microscopic expression of $\chi_{\rm em}$. 
Namely, the four susceptibilities can be put together to form the 
single susceptibility $\chi_{\rm em}$.  In this sense, "the Casimir 
form derived from (D)", to be called scheme (A)*, is a single 
susceptibility theory.  But the Casimir form 
with phenomenologically determined susceptibility, scheme (A), has 
no guarantee to be a single susceptibility theory.    \\

The argument given above shows that the four schemes can be transformed 
to one another.  From an arbitrary constituive equation one can derive 
all the other ones.  This means that the dispersion equation should be 
same for all the schemes, i.e., the one already known in (D) 
\begin{equation}
 {\rm det} |[k^2 - (\frac{\omega}{c})^2] {\bf 1} 
     - \mu_{0} \chi_{\rm em}^{(\rm T)}(\Vec{k},\omega)| = 0  
\end{equation} 
can be used also for (A)*, (B), and (C). \\

To sum up, the scheme (D) is conceptually the simplest and 
practically the most informative scheme at present among the possible 
forms of macroscopic M-eqs. \\

The author acknowledges the discussions with Dr. A. Chipouline and 
S. Tretyakov for improving his understandings about the schemes 
(A), (B), and (C). 


\end{document}